\documentclass[%
 reprint,
 superscriptaddress,
 amsmath,amssymb,
 aps,
floatfix,
]{revtex4-2}

\usepackage{graphicx}
\usepackage{dcolumn}
\usepackage{bm}
\usepackage{color}
\usepackage{xcolor}
\usepackage{hyperref}
\usepackage[dvipsnames]{xcolor}
\usepackage{mhchem}

\begin{document}

\title{Direct observation of vortex liquid droplets in the iron pnictide superconductor {CaKAs$_4$Fe$_4$} at 0.5T$_c$}

\author{Óscar Bou Marqués}
\affiliation{Laboratorio de Bajas
Temperaturas, Departamento de Física de la Materia Condensada, Instituto Nicolás Cabrera and Condensed Matter Physics Center (IFIMAC), Unidad Asociada
UAM-CSIC, Universidad Autónoma de Madrid, E-28049 Madrid, Spain.
}

\author{Jose A. Moreno}
\affiliation{Laboratorio de Bajas
Temperaturas, Departamento de Física de la Materia Condensada, Instituto Nicolás Cabrera and Condensed Matter Physics Center (IFIMAC), Unidad Asociada
UAM-CSIC, Universidad Autónoma de Madrid, E-28049 Madrid, Spain.
}

\author{Pablo Garc\'ia Talavera}
\affiliation{Laboratorio de Bajas
Temperaturas, Departamento de Física de la Materia Condensada, Instituto Nicolás Cabrera and Condensed Matter Physics Center (IFIMAC), Unidad Asociada
UAM-CSIC, Universidad Autónoma de Madrid, E-28049 Madrid, Spain.
}

\author{ Mingyu Xu}
\altaffiliation{Currently at: Department of Chemistry, Michigan State University, East Lansing, Michigan 48824, USA}
\affiliation{Ames National Laboratory, Iowa State University, Ames, Iowa 50011, USA}
\affiliation{
Department of Physics and Astronomy, Iowa State University, Ames, Iowa 50011, USA
}

\author{Juan Schmidt}
\affiliation{Ames National Laboratory, Iowa State University, Ames, Iowa 50011, USA}
\affiliation{
Department of Physics and Astronomy, Iowa State University, Ames, Iowa 50011, USA
}
\affiliation{Departamento de Física, FCEyN, Universidad de Buenos Aires, Buenos Aires 1428, Argentina
}

\author{Sergey L. Bud'ko}
\affiliation{Ames National Laboratory, Iowa State University, Ames, Iowa 50011, USA}
\affiliation{
Department of Physics and Astronomy, Iowa State University, Ames, Iowa 50011, USA
}
\author{Paul C. Canfield}
\affiliation{Ames National Laboratory, Iowa State University, Ames, Iowa 50011, USA}
\affiliation{
Department of Physics and Astronomy, Iowa State University, Ames, Iowa 50011, USA
}
\author{Isabel Guillamón}
\affiliation{Laboratorio de Bajas
Temperaturas, Departamento de Física de la Materia Condensada, Instituto Nicolás Cabrera and Condensed Matter Physics Center (IFIMAC), Unidad Asociada
UAM-CSIC, Universidad Autónoma de Madrid, E-28049 Madrid, Spain.
}
\author{Edwin Herrera}
\affiliation{Laboratorio de Bajas
Temperaturas, Departamento de Física de la Materia Condensada, Instituto Nicolás Cabrera and Condensed Matter Physics Center (IFIMAC), Unidad Asociada
UAM-CSIC, Universidad Autónoma de Madrid, E-28049 Madrid, Spain.
}
\author{Hermann Suderow}
\affiliation{Laboratorio de Bajas
Temperaturas, Departamento de Física de la Materia Condensada, Instituto Nicolás Cabrera and Condensed Matter Physics Center (IFIMAC), Unidad Asociada
UAM-CSIC, Universidad Autónoma de Madrid, E-28049 Madrid, Spain.
}







\begin{abstract}
Type-II superconductors under magnetic fields are in a quantum coherent non-dissipative state as long as vortices remain pinned. Dissipation appears when vortices depin, eventually driven by thermal fluctuations. This can be associated to a melting transition between a vortex solid and a vortex liquid. This transition is almost always observed very close to T$_c$ when probed by macroscopic experiments. However, it remains unclear how the vortex solid responds to thermal fluctuations at the scale of individual vortices far from the melting transition. Here we use scanning tunneling microscopy (STM) to visualize vortices in CaKAs$_4$Fe$_4$ (T$_c \approx$ 35 K). We find vortex liquid droplets---localized regions in space where vortices strongly fluctuate due to thermal exctiation---at temperatures as low as 0.5\,T$_c$. Our results show that the onset of dissipation at the local scale occurs at temperatures considerably below T$_c$ in type-II superconductors.
\end{abstract}

\maketitle

\section*{Introduction}
In type II superconductors the magnetic field can penetrate in the material in the form of superconducting vortices. Inside the vortex core the superconducting pair wavefunction is suppressed on the scale of the superconducting coherence length $\xi$. In a clean superconductor vortices arrange themselves in an ordered lattice\,\cite{abrikosov1955magnetic}. But real materials rarely maintain this ordered ideal vortex solid. Vortex cores threading regions with crystalline defects, for example, have a smaller core free energy and such vortices tend to remain pinned to the defected regions. This, combined with the variations in the intervortex interactions as a function of temperature and magnetic field and with thermal fluctuations, eventually leads to a phase transition between a vortex solid and a vortex-liquid\,\cite{blatter1994vortices, brandt1995flux,nelson1993boson,Klein2001,PhysRevB.68.054509,PhysRevB.69.024501}. Vortices are essentially mobile in the vortex liquid phase, which may produce dissipation under the application of an infinitesimal perturbation as for example, a small current\,\cite{10.1063/1.1401180,Campbell01031972,PhysRevLett.58.599,PhysRevB.62.671}. 

The relevance of thermal fluctuations for a certain superconductor can be described by the Ginzburg-Levanyuk number $G_i$\,\cite{ginzburg1961some,levanyuk1959contribution}. $G_i$ compares the critical temperature $T_c$ with the free energy of the superconductor in a volume defined by the coherence length $\xi$, $G_i=\frac{k_B T_c}{H_c^2 \xi^3}$ (for isotropic superconductors, where $H_c$ is the thermodynamic critical field and $k_B$ Boltzman's constant). $G_i$ is directly related to thermally induced motion of vortices, which is more relevant for higher values of $G_i$\,\cite{blatter1994vortices,brandt1995flux,Eley2017,PhysRevB.64.184514,PhysRevB.69.024501}. $G_i$ can be also related to the temperature and field range presenting a vortex liquid phase\,\cite{koshelev2019melting}. The vortex liquid phase is practically absent in low T$_c$ superconductors with $G_i\approx10^{-7}$ and occurs in a temperature range a few \% below T$_c$ in cuprate superconductors with $G_i\approx 10^{-3}-10^{-2}$.

The melting process of the vortex solid in high magnetic fields, where the intervortex separation is larger but of the order of $\xi$, has been studied by macroscopic measurements including specific heat or magnetization\,\cite{zeldov1995thermodynamic, safar1992experimental, willemin1998first,Paltiel2000,PhysRevLett.78.4833}, microscopic but spatially averaged techniques as small angle neutron scattering\, \cite{Cubitt1993,f945bee77a6d41b083b4f8003e5ec2a3,Yaron1995}, and through direct visualization in real space such as scanning tunneling microscopy (STM)\,\cite{PhysRevLett.62.214,RevModPhys.79.353,Suderow2014,Guillamon2009,Guillamon2014,Duhan2025,PhysRevLett.122.047001,PhysRevLett.106.077001,PhysRevLett.70.505,PhysRevLett.65.2583,PhysRevLett.91.127002,PhysRevB.57.6061,Hecher2014,Zehetmayer2015,PhysRevLett.103.257001}. Other spatially resolved techniques as Magnetic Force Microscopy mostly address the range of small magnetic fields and intervortex separations much larger than $\xi$\,\cite{Embon2017}. STM allows to image individual vortices in the field range which is mostly used in applications\,\cite{10.1063/1.1401180,Campbell01031972,10.1063/1.4874979}. Often, the electronic density of states at the Fermi level inside vortex cores is much larger than the density of states outside vortex cores (the latter is, for low fields, similar to the one obtained in zero field superconducting density of states measurements). By making maps of the zero bias tunneling conductance, which is proportional to the Fermi level electronic density of states, many individual vortices are imaged in a large range of magnetic fields\,\cite{PhysRevLett.62.214,RevModPhys.79.353,Suderow2014,Guillamon2009,Guillamon2014,Duhan2025,PhysRevLett.122.047001,PhysRevLett.106.077001,PhysRevLett.70.505,PhysRevLett.65.2583,PhysRevLett.91.127002,PhysRevB.57.6061,Hecher2014,Zehetmayer2015}. However, vortex studies with STM have been mostly restricted to superconductors with very small $G_i$.  Measurements of the vortex phases as a function of temperature in high T$_c$ cuprate superconductors with large $G_i$ show mostly vortex solid phases with different degrees of thermally induced disorder\,\cite{PhysRevB.72.014525}. The observation of melting remains challenging, because the density of states in and around vortices is intricate and difficult to follow with temperature\,\cite{PhysRevLett.85.1536,doi:10.7566/JPSJ.82.083706,doi:10.1126/science.aat1773,PhysRevLett.75.2754}.

\begin{figure}[htb]
\includegraphics[width=\columnwidth]{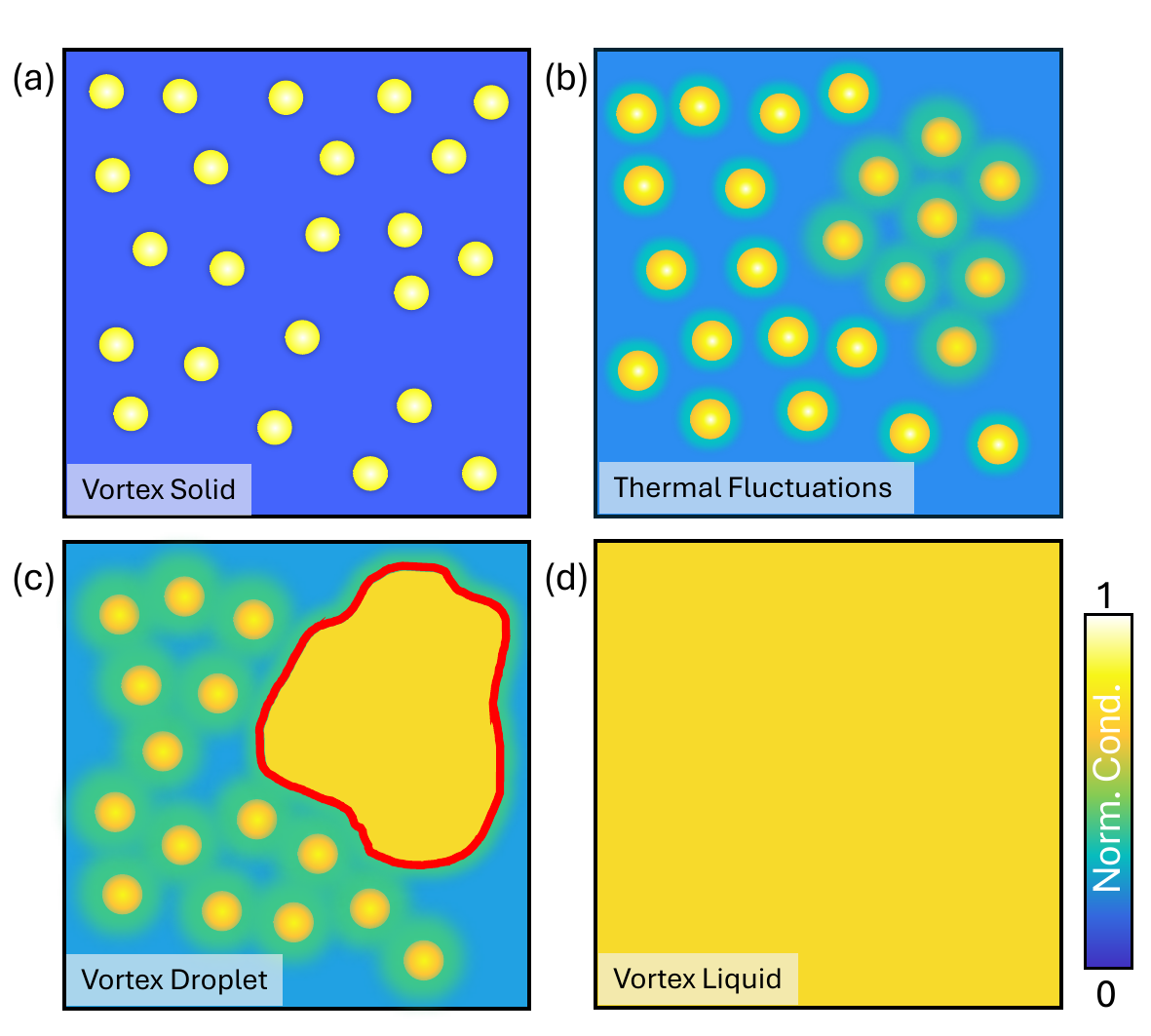}
\caption{{\bf Schematic representation of vortex solid and liquid phases.} (a) At low temperatures, the vortices are pinned and eventually form a disordered solid. The superconducting state (dark blue) is clearly distinguished with respect to the normal state at the center of the vortex core (white). (b) As the temperature rises, vortices begin to move, hopping between different positions. Vortex motion leads to fluctuations in the vortex positions which are much faster than our measurement time, resulting in an apparent increase in the vortex core size (the time-averaged density of states is represented by the green halo). (c) In certain areas we eventually observe ``vortex liquid droplets'' (large yellow area marked by the red line). These droplets coexist with a vortex solid. (d) Close to the critical temperature, $T_c$, of the superconductor, we observe a vortex liquid spanning large fractions of the field of view. The vortex liquid phase consists of large areas with a spatially uniform tunneling conductance which is below the tunneling conductance at high values, evidencing the presence of a superconducting gap.}
\label{scheme}
\end{figure}

Owing to their relatively large T$_c$ and small coherence length $\xi$, iron pnictide superconductors provide enhanced fluctuations and a relatively large vortex liquid range in the temperature-magnetic field phase diagram\,\cite{koshelev2019melting,Eley2017}. Critical temperatures as high as T$_c=$ 35 K occur in the stoichiometric material CaKFe$_4$As$_4$, which also presents a coherence length $\xi_{ab}$ below 2 nm\,\cite{doi:10.1021/jacs.5b12571,meier2016anisotropic,PhysRevB.95.100502,Canfieldphase}. With $G_i\approx $ 10$^{-4}$, a well defined density of states inside vortex cores\cite{PhysRevB.97.134501}, and clear a-priori knowledge of most relevant pinning centers (CaFe$_2$As$_2$ and KFe$_2$As$_2$ intergrown layers \cite{Ichinose2021,He2025,he2025defect,Takahashi2020,Kobayashi2020,Haberkorn2020,Haberkorn2019,Chen2025,huyan2025,cui2017magnetic}), CaKFe$_4$As$_4$ arises as an excellent system to study thermal fluctuations at the scale of individual vortices. Here we address the vortex solid and liquid phases of CaKFe$_4$As$_4$ using STM and unveil vortex liquid droplets appearing locally well below T$_c$.

It is important to realize that STM measures at time scales of several tens of minutes. In the presence of thermal fluctuations, the tunneling conductance is acquired at a single point and is the result of the time average over a vortex changing its position continuously below the tip. The tip rests on top of a vortex generally for a few seconds (depending on the vortex density and image size). Except at locations with exceptionally slow dynamics (discussed below), taking tunneling conductance measurements as a function of time at a single vortex generally does not provide a significant time dependent signal, suggesting that the time scale for most fluctuations is well above the fastest measurement time scale, determined by the acquisition bandwidth (of about 10 kHz)\,\cite{10.1063/1.1939077}. If the amplitude of fluctuations is no bigger than the vortex core, we see perfectly defined cores just as shown in Fig.\,\ref{scheme}(a). When fluctuations are larger than the vortex core size, the tunneling conductance map shows an apparent increase in the core size, as schematically represented in Fig.\,\ref{scheme}(b) and in the portion of (c) showing the vortex solid. We identify the vortex liquid when we do not observe individual vortices at locations where we observe individual vortices at lower temperatures. We consider that areas presenting a spatially uniform zero bias tunneling conductance and which are larger than three times the average intervortex distance are in the vortex liquid state. In the vortex liquid, Fig.\,\ref{scheme} (d) and the portion of (c) showing the vortex liquid, we observe a tunneling conductance at zero bias below its value at high bias (schematically shown by the yellow color, instead of white for the normal phase, in Fig.\,\ref{scheme})\,\cite{Guillamon2009,Guillamon2014,Duhan2025,PhysRevLett.122.047001}.

\begin{figure*}[htb]
\includegraphics[width=\textwidth]{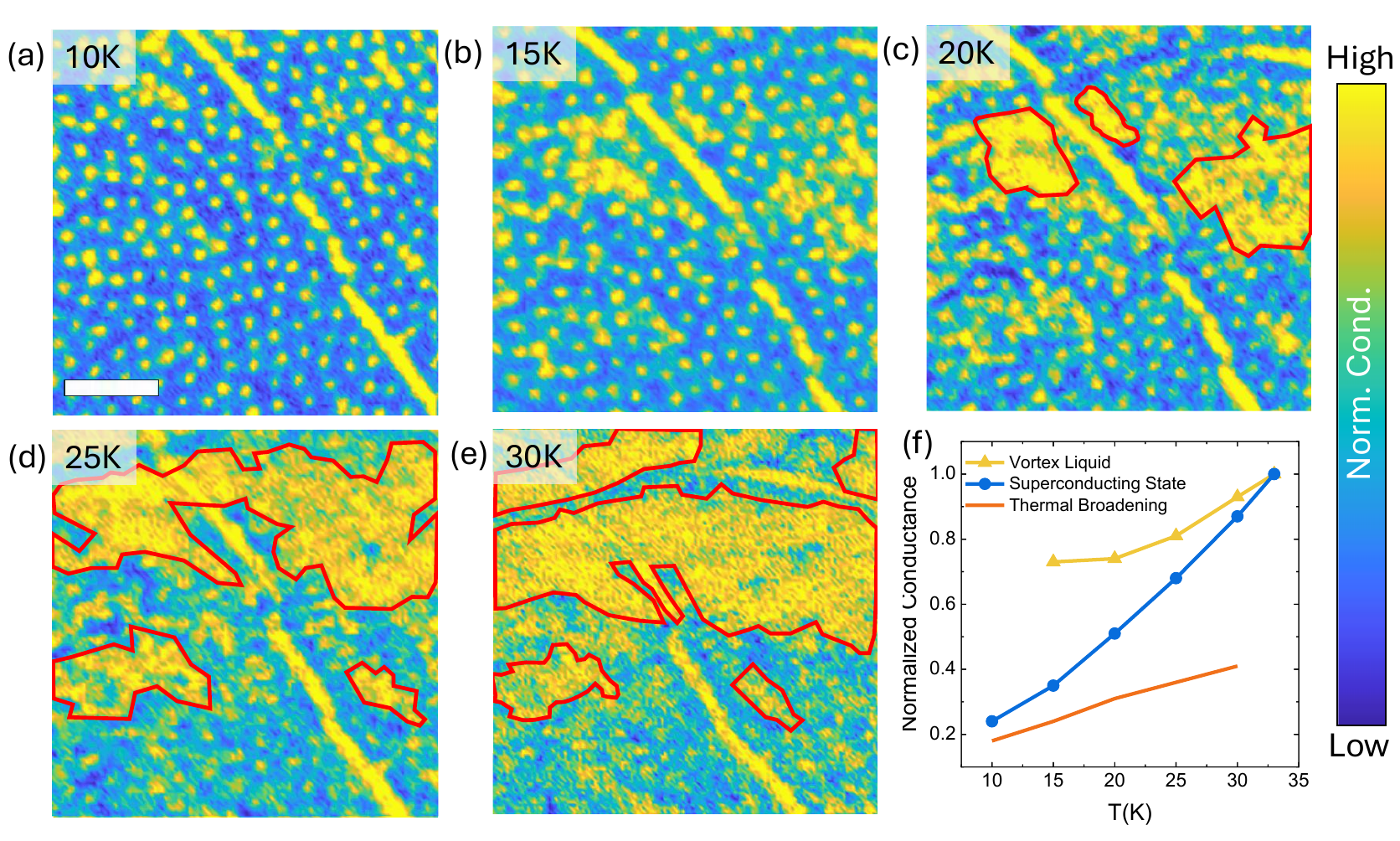}
\caption{{\bf Direct observation of vortex liquid droplets well below T$_c$ in CaKAs$_{4}$Fe$_4$.} In (a-e) we show maps of the zero bias tunneling conductance for different temperatures and at 10 \,T, in approximately the same field of view. Linear defects found in the surface of the sample lead to the roughly diagonal yellow lines shown in the maps (more information in the Supplementary Information Section I and Supplementary Figures\,1. and 2; maps at other temperatures and magnetic fields are commented and provided in the Supplementary Information Section III and in the Supplementary Figure\,4). At higher temperatures we mark with red contours the areas where we do no longer identify isolated vortices in the tunneling conductance maps. In these areas, the tunneling conductance is still well below that of the normal phase. The white scale bar in panel (a) corresponds to 40 nm. The color scale corresponds to slightly different conductance values in each figure, to maximize contrast. To discuss the value of the conductance in (a-e) we show in (f) the tunneling conductance at zero bias normalized to its high bias value versus temperature, taken at different positions. Lines connect points and are guides to the eye. We show the tunneling conductance taken in between vortices (blue areas in (a-e)) as blue circles. The change induced by the temperature smearing in the zero bias tunneling conductance is shown as an orange line. We also show as yellow triangles the tunneling conductance inside the vortex liquid droplets (average over areas within the red lines in a-e).}
\label{melting}
\end{figure*}

\section*{Results}

In Fig.\,\ref{melting}(a-e), we show a characteristic example of zero bias conductance maps at different temperatures within the same field of view. At 10\,K (Fig.\,\ref{melting}(a)) we observe individual vortices all over the image. Intervortex distance follows the prediction for the triangular Abrikosov vortex lattice, as shown at lower temperatures in Ref.\,\cite{PhysRevB.97.134501}. There are also features in the vortex maps where the tunneling conductance forms lines with a conductance slightly below, but close to the normal phase value. These linear features can be associated with lines observed in the STM topography, see Supplementary Information Section I and Supplementary Fig.\,1. At 15\,K and at higher temperatures (Fig.\,\ref{melting}(b-e)) we observe that there are areas, which we call vortex liquid droplets, where vortices merge into increasingly larger patches, within the areas marked by a red line. These areas coexist with the disordered vortex solid, where we can still resolve individual vortices. Vortex liquid droplets are observed at temperatures as small as 0.5\,T$_c$. At the highest temperature, 30\,K (T$_c$ at this field is of about 32\,K as obtained from resistivity\,\cite{Canfieldphase}), vortex liquid areas encompass nearly half the whole field of view.

It is important to plot the normalized tunneling conductance in different parts of the field of view as a function of the temperature (Fig.\,\ref{melting}(f)). The tunneling conductance $\sigma(V)$ is given by $\sigma(V)=\int dE N(E) \frac{df(E-eV)}{dE}$, where $N(E)$ is the density of states, $f(E)$ the Fermi function and E the energy. $N(E)$ for CaKFe$_4$As$_4$ has been discussed before and is understood as a modified BCS density of states described in Refs.\,\cite{PhysRevB.95.100502,PhysRevB.97.134501,PhysRevLett.117.277001}. The zero energy density of states is zero, except on impurities or defects\,\cite{PhysRevB.95.100502,PhysRevB.97.134501}. With increasing temperature, the superconducting gap decreases, and there is an enhanced $\sigma(V=0)$ due to the temperature smearing by the Fermi function. Its derivative, $\frac{df(E-eV)}{dE}$, changes from close to a $\delta$ function at absolute zero to a smeared peak with increasing temperature. The temperature induced increase in the zero bias tunneling conductance $\sigma(V)$ is shown by the orange line in Fig.\,\ref{melting}(f). This corresponds to the temperature induced reduction of the contrast in observing vortices by measuring the tunneling conductance. We see however that the temperature variation of the zero bias tunneling conductance in between vortices (blue circles in Fig.\,\ref{melting}(f)) is much stronger. This is due to thermally induced vortex core overlap. We discuss this with more detail in the Supplementary Information Section II and Supplementary Figure\,3 (maps taken at different temperatures and magnetic fields are shown in the Supplementary Information Section III and in the Supplementary Figure\,4). Within the vortex liquid droplets, the tunneling conductance at zero bias is very different from the one in between vortices and also increases with temperature (yellow triangles in Fig.\,\ref{melting}(f)). The zero bias tunneling conductance remains always below the normal phase value, evidencing the motion of vortices inside the liquid droplets.

\begin{figure*}[htb]
\includegraphics[width=\textwidth]{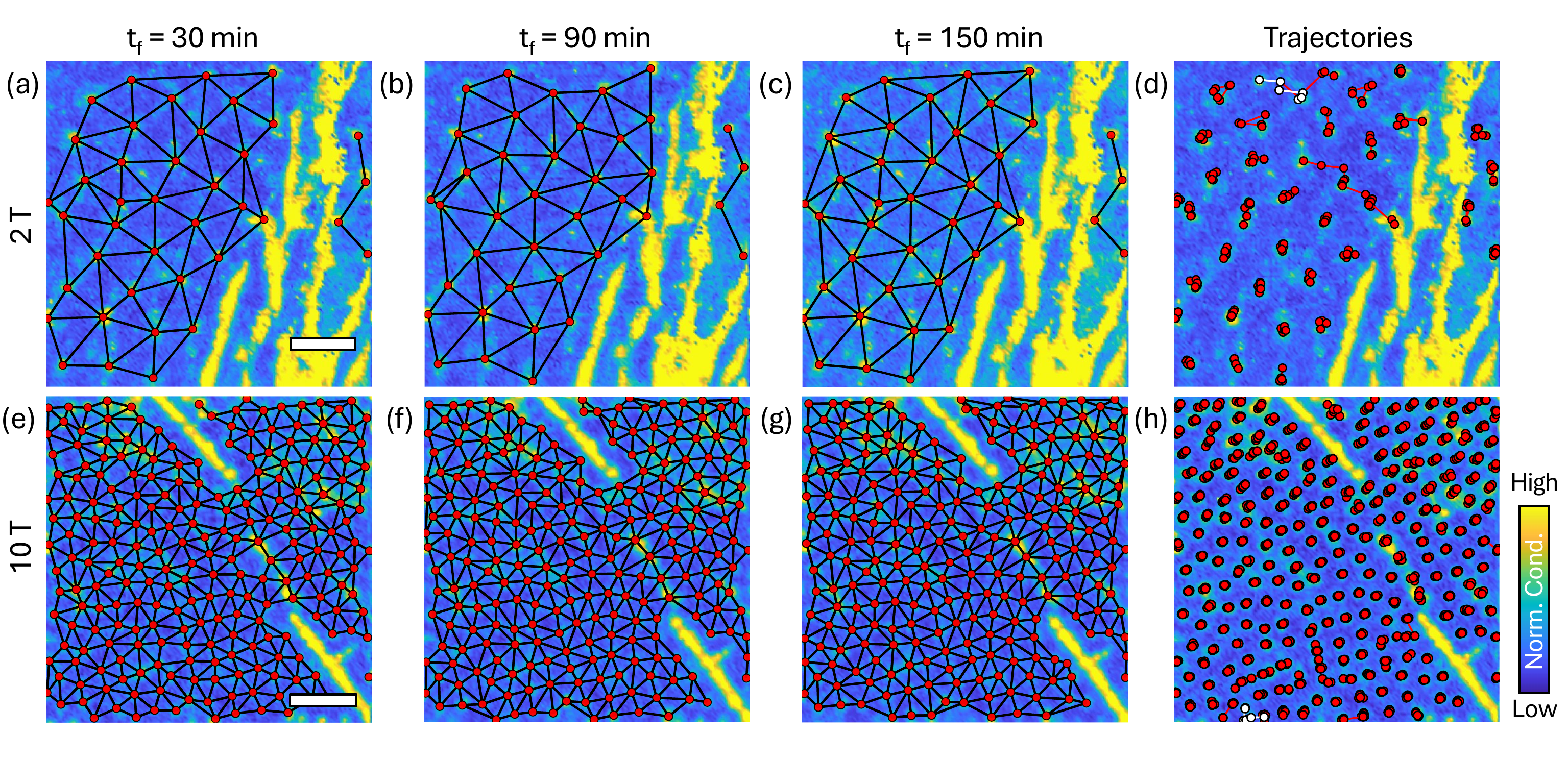}
\caption{{\bf Thermal fluctuations of vortices.} In (a-c) and in (e-g) we show zero bias tunneling conductance maps acquired sequentally at, respectively, 2\,T and 10\,T. Each map has been taken in half an hour, and the field of view remains the same at each magnetic field. Vortex positions are marked by red circles. The black lines join nearest neighbour vortices, obtained by triangulation. In (d,h) we superimpose red circles united by a red line for each magnetic field ((d) for 2\,T and (h) for 10\,T) on the first zero bias tunneling conductance map of the sequence. These circles and red lines mark the trajectory of each vortex in the sequence. We see that most vortices remain at the same position. However, a few vortices have moved. For example, the vortices marked in white in the top central part of (d), or at the bottom center in (h), are the vortices in each case that present the largest displacement in the field of view. Note that the largest displacement is much larger at 2 T than at 10 T. In total, we acquired six consecutive images at each magnetic field. White bars in (a) and (e) correspond to 40 nm.}
\label{Creep}
\end{figure*}

We note that the liquid vortex droplets are formed at low temperatures in areas which are well separated from the linear defects (Fig.\,\ref{melting}(b)). At higher temperatures (Fig.\,\ref{melting}(e)), the vortex liquid droplets join the linear features in the tunneling conductance maps. We find that vortex liquid droplets preferentially form at areas where vortices present enhanced fluctuations at low temperatures.

\begin{figure}[htb]
\includegraphics[width=1\columnwidth]{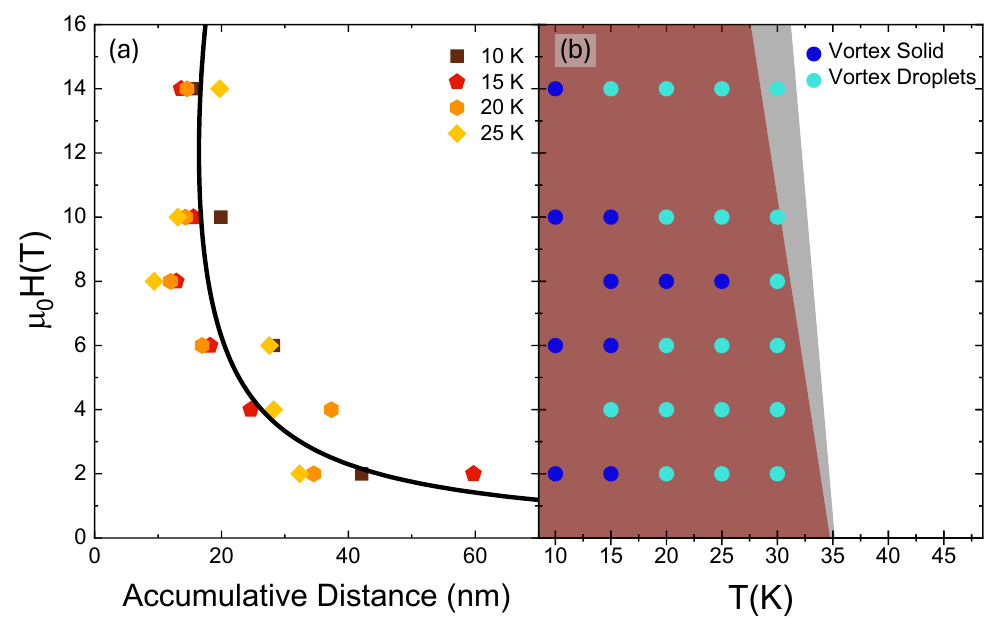}
\caption{{\bf Vortex melting phase diagram obtained from local measurements.} In (a) we show the distance traveled by vortices through thermal motion as a function of the magnetic field, for the temperatures provided in the legend. These data are taken inside areas with a well defined vortex solid. The black line corresponds to the displacement calculated using the vortex pinning force density $f_{n}$ given by the Dew-Hughes model\,\cite{dew1974pinning} with parameters  $H_{c2} = 50$ T, $p = 0.95$ and $q = 3$. In (b), we show the phase diagram from macroscopic measurements, obtained from Refs.\,\cite{Canfieldphase,Cheng2019}, with grey and maroon areas representing vortex solid and vortex liquid phases. We show as circles the temperatures and fields that we have explored. Dark blue circles signal areas where we observe a vortex solid, resolving individual vortices over nearly the whole field of view. With cyan circles, we show temperatures and fields where we have found vortex liquid droplets or areas with strongly fluctuating vortices.}
\label{dynamics_phase_diagram}
\end{figure}

To explore the role of pinning in the melting process in more detail, we now discuss zero bias conductance maps taken consecutively at a fixed temperature and magnetic field. In Fig.\,\ref{Creep}, we show several frames taken at 10\,K, and at 2\,T (Fig.\,\ref{Creep}(a-d)) and 10\,T (Fig.\,\ref{Creep}(e-h)). We show as red circles the positions for all vortices in all frames. All acquired maps are discussed and presented in the Supplementary Information Section IV and the Supplementary Figures\,5-10. The vortex nearest neighbors are found following Delaunay triangulation and shown by the black lines in Fig.\,\ref{Creep}(a-c,e-g). We first remark that most vortices stay at roughly the same position in the different tunneling conductance maps. At 2\,T (Fig.\,\ref{Creep}(a-c)), there are however several vortices which move considerable distances.

As we show in Fig.\,\ref{Creep}(d), at 2 T, these vortices travel distances far above the average intervortex distance. We have calculated the accumulated maximal distance traveled by vortices within the time frame of Fig.\,\ref{Creep} ($\sim $150 minutes) at each magnetic field and temperature. The result is shown as points in Fig.\,\ref{dynamics_phase_diagram}(a). We can see that low magnetic fields allow for a strong vortex mobility by thermal fluctuations. Around 8 T and 10 T the vortex presenting largest mobility travels less than the intervortex distance (18 nm for 8 T and 16 nm for 10 T). It is relevant to note that data for all temperatures roughly fall into a single curve.

We can now compare with macroscopic measurements (Fig.\,\ref{dynamics_phase_diagram}(b)). Maroon and grey shaded temperatures and magnetic fields represent vortex solid and liquid phases as obtained from macroscopic measurements from Refs \cite{Canfieldphase,Cheng2019}. Points show the temperatures and magnetic fields which we have explored. We see that the temperature and field range where we observe vortex liquid droplets goes down to 0.5\,T$_c$. Furthermore, we also remark that there can be a certain relationship between thermally induced vortex mobility Fig.\,\ref{dynamics_phase_diagram}(a) and the lowest temperature at which we observe liquid vortex droplets. When the vortex mobility is smallest at 8\,T, the temperature for the appearance of liquid vortex droplets tends to be closer to T$_c$.

We note that vortex fluctuations are not always occuring at time scales smaller than our measurement time. In the Supplementary Information Section V we discuss an extreme case of slow vortex fluctuation, where a vortex is observed to jump between two different positions which evolve with time at the scale of hours. Generally, our data are taken with increasing temperature. We do not observe temperature induced hysteresis phenomena. Although this point probably requires more measurements, out of the scope of present work, it is not very surprising given the large time scales needed for our measurements.

\section*{Discussion}

The vortex liquid droplets observed here occur at temperatures well below the temperature where the vortex liquid is identified from macroscopic magnetization\,\cite{Canfieldphase,meier2016anisotropic,Wang2020vortex,Wang2020vortex2,Cheng2019}, and magnetic force microscopy measurements of individual vortices at small magnetic fields (up to 0.008\,T)\,\cite{He2025}. On the other hand, transport and magnetization measurements have found ultrahigh critical current densities in single crystals of CaKFe$_4$As$_4$\,\cite{Canfieldphase,Haberkorn2020}. Uniquely to this material, critical currents have shown a non-monotonic temperature and magnetic dependence, i.e, critical currents at 20\,K exceed those at 5\,K at high magnetic fields ($\gtrsim$2\,T)\cite{Ishida2019,Pyon2019}. This effect is more pronounced than previously observed ``fish-tail'' magnetization curves in cuprates\cite{Daeumling1990} and other iron-based superconductors\cite{Yang2008,Taen2009,PhysRevB.78.224506,PhysRevB.81.094509,PhysRevB.84.060509,PROZOROV2009667} and suggests a distinctive vortex pinning situation. Along the same line, planar defects, CaFe$_2$As$_2$ and KFe$_2$As$_2$ intergrown layers, are the main vortex pinning centers (see also Supplementary Information Section I and Supplementary Fig. 2)\,\cite{Ishida2019,Pyon2019,Sugali2021,Ichinose2021,He2025,he2025defect,Takahashi2020,Kobayashi2020,Haberkorn2020,Haberkorn2019,Chen2025}.

By following individual vortices we obtain that the maximal distance traveled by vortices from thermal fluctuations (Fig.\,\ref{dynamics_phase_diagram}(a)) presents a curve which is nearly independent of temperature. This curve reminds us of the ones obtained using the Dew-Hughes model from Ref.\,\cite{dew1974pinning}. Within this model, the interplay between the elastic intervortex interaction and the pinning landscape leads to a magnetic field dependence of the pinning force which falls into a single curve for all temperatures and is of the form $\propto h^{p}(1-h)^{q}$ with $h = \frac{H}{H_{c2}}$ and $p$ and $q$ being two fitting parameters. Our data follow this field dependence, as shown by the black line in Fig.\,\ref{dynamics_phase_diagram}(a), using $H_{c2} = 50$ T, $p = 0.95$ and $q = 3$. Assuming that the maximal distance traveled we determine here is directly related to the pinning force, we find that the field where the force is maximal (the traveled distance minimal) is of $h_{max} \approx$ 0.24, similar to the value of $h_{max} \approx$ 0.28 found in Ref.\,\cite{Canfieldphase,Haberkorn2020cafeas} for single crystals. The model predicts a pinning mechanism governed by surface pinning for normal center of core interaction for $h_{max}=0.2$, and by point cores for $h_{max}=0.33$\,\cite{dew1974pinning}. We expect that the CaFe$_2$As$_2$ and KFe$_2$As$_2$ intergrowth layers act as these surface-like pinning centers. The scaling of the observed traveled distance is quite good (all data points falling into a single dependence in Fig.\,\ref{dynamics_phase_diagram}(a)) but it is not perfect either. This suggests that a single mechanism does not govern the pinning; point-like disorder and pinning induced by the large linear defects (see Supplementary Information Section I) possibly coexist and have both a strong influence in vortex positions.
We also find that, independently of the pinning centers, the minimum motion of the vortices corresponds directly with the place in the phase diagram (Fig.\,\ref{dynamics_phase_diagram}(b)) where the vortex solid is more resistant to the formation of local vortex liquid droplets. This illustrates how the pinning potential landscape governs the motion of vortices and the formation of vortex liquid droplets alike.

We note that the vortex lattice in a material with a strong four-fold anisotropy in the normal density of states, as CaKFe$_4$As$_4$, might show at sufficiently high magnetic fields a transition between hexagonal and square ordered vortex lattices\,\cite{PhysRevLett.79.487,PhysRevB.79.174522,yaron1996microscopic}. It seems that vortices are too disordered in present samples of CaKFe$_4$As$_4$ to establish such a phase transition over macroscopic length scales\,\cite{PhysRevB.97.134501}. However, elastic moduli are strongly influenced by electronic interactions and some moduli, as the rotation modulus, can vanish at certain conditions\,\cite{PhysRevLett.87.137002}. This feature could favor the formation of places with strong vortex fluctuations.

Our data raise the question about the observation of a finite voltage in macroscopic critical current measurements. Possibly, even in presence of large amounts of vortex liquid droplets, sufficiently large samples still present zero voltage drop, owing to the formation of percolation paths for current flow\,\cite{wu2024globalcriticalcurrenteffect}. Nevertheless, small micron sized samples, as those used in research of two-dimensional materials and in nanostructured superconductors, should be much more sensitive to the formation of vortex liquid droplets\,\cite{doi:10.1126/science.adv8376,Navarro-Moratalla2016,Fridman2025,Cordoba2020}.

We note that up-to-date measurements of the melting transition in two-dimensional vortex solids, obtained in thin films with a perpendicular magnetic field, present  a melting transition where fluctuations induce defects in an otherwise ordered hexagonal lattice\,\cite{Guillamon2009,Guillamon2014,Duhan2025,PhysRevLett.70.505,PhysRevLett.65.2583,PhysRevLett.91.127002}. The transition to a vortex liquid where individual vortices are no longer resolved with STM occurs over large portions of the sample. The observation of vortex liquid droplets here shows a distinct pinning landscape consisting of a combination of crystal lattice and defects in CaKFe$_4$As$_4$.

\section*{Conclusion}

In summary, we have imaged superconducting vortices as a function of temperature and magnetic field in the iron based superconductor CaKAs$_4$Fe$_4$. We have observed a disordered vortex solid at low temperatures which melts locally, forming vortex liquid droplets well below the vortex melting temperature. Thus, the melting of vortices in superconductors is not a uniform transition, but is strongly linked to the local pinning potential. We have explored this link directly, by characterizing thermally activated vortex motion and finding a close relationship between vortex pinning and thermally induced motion. Our data reveal a broad crossover region where solid, and liquid vortex phases coexist. This suggests that the actual dissipationless superconducting phase has a much smaller temperature range than what can be found from macroscopic experiments, underscoring the fundamental relevance of studies aimed to enhance vortex pinning for building functional superconducting devices.

\section*{Acknowledgements}

This work was supported by the Spanish Research State Agency (TED2021-130546B\-I00, PID2023-150148OB-I00 and CEX2023001316-M), Comunidad de Madrid through project TEC-2024/TEC-380 “Mag4TIC”, the EU through grant agreement No 871106 and by the European Research Council PNICTEYES grant agreement 679080 and VECTORFIELDIMAGING grant agreement 101069239. We acknowledge the QUASURF project (SI4/PJI/2024-00199) funded by the Comunidad de Madrid through the agreement to promote and encourage research and technology transfer at the Universidad Aut\'onoma de Madrid. We acknowledge collaborations through EU program Cost CA21144 (www.superqumap.eu). Work at Ames National Laboratory (crystal growth, basic characterization, drafting and editing) was supported by the U.S. Department of Energy, Office of Basic Energy Science, Division of Materials Sciences and Engineering. Ames National Laboratory is operated for the U.S. Department of Energy by Iowa State University under Contract No. DE-AC02-07CH11358. We also acknowledge SEGAINVEX for support in design and for construction of the temperature stabilized STM.

\section*{Methods}

Single crystals of CaKAs$_4$Fe$_4$ were grown and characterized following the methods described in \cite{meier2016anisotropic, Meier2017}. We cleaved the samples by glueing a post to the top of each sample and pushing it with the slider described in Ref.\,\cite{Descmicroscopio}. With this method, we obtained large ($>2\mu$m $\times$ \,2$\mu$m), clean and atomically flat surfaces. The tip was made of gold and was conditioned (cleaning and sharpening) using a gold sample and following the methods described in Ref.\,\cite{SharpeningAu}. We used a modified lab-built STM system, similar to the one described in Ref.\,\cite{Descmicroscopio}. The control software is described in Ref.\,\cite{Descsoftware} and available in Ref.\,\cite{githublbt}. Image rendering was performed mostly using the lab-built software described in Ref.\,\cite{Descsoftware}, using occasionally imaging software as Ref.\,\cite{10.1063/1.2432410}. The system was introduced in a helium bath cryostat at 4.2\,K on a support that was thermally anchored to a cold plate, itself thermally anchored to the helium bath\,\cite{Montoya2019}. We used a heater and a temperature controller to modify the temperature of the STM system. We reduced thermal creep from temperature gradients by placing the heater between the cold source and the STM holder. We maintained temperature stability of $\pm$ 100\,mK up to temperatures above 40\,K. We tested thermal stability by making subsequent topographic STM maps and verifying that the field of view remained the same. When required, we modified the field of view using references in the topography. To map vortices, we acquire the tunneling conductance as a function of the bias voltage. Maps are generally made by normalizing the tunneling conductance to a value at a bias voltage well above the superconducting gap (above 10 mV) and plotting as a function of the position the zero bias tunneling conductance. As the temperature variation of the tunneling conductance is quite pronounced, we maximize the color scale at each temperature (unless otherwise stated) to best visualize vortices. The superconducting state is represented as blue, and the normal state inside the vortex as yellow in the maps shown below. At each applied magnetic field we have increased the temperature in steps of 5\,K between 10\,K and 30\,K. We also acquired several maps, one after the other at each temperature, to follow the  positions of individual vortices. Each maps was taken within 30 minutes. Delaunay triangulation of the vortex positions was performed using the software available in Ref.\,\cite{githublbt} and described in Refs.\,\cite{Guillamon2014,PhysRevResearch.2.013329,PhysRevB.95.134505,PhysRevResearch.2.013125}. Using the in-situ positioning method described in Ref.\,\cite{Descmicroscopio}, we explored many different fields of view, in regions with different numbers of defects in two different samples.

\section*{Data Availability}

The data shown in this study is available from the corresponding author upon reasonable request.

\section*{Code Availability}

The code used to analyse this results can be found in \cite{githublbt}.

\section*{Author Contributions}

OBM and JAM performed the experiments, with the supervision of EH, IG and HS. The analysis of the data and the coding of the tools used to do so were made by OBM and PGT. Samples were synthesized and characterized by MX, JS, SLB and PCC. The paper was written by OBM, JAM, IG, EH, and HS with input and approval from all authors.

\section*{Competing Interests}

The authors declare no competing interests.

\section*{Corresponding author}

Corresponding author: hermann.suderow@uam.es.

\clearpage

\section*{Supplementary Information}

\section{Surface Characterization and defects}

\renewcommand{\thefigure}{\arabic{figure}}

\renewcommand{\figurename}{Supplementary Figure} 
\setcounter{figure}{0}

\begin{figure*}[htb]
\includegraphics[width=1\textwidth]{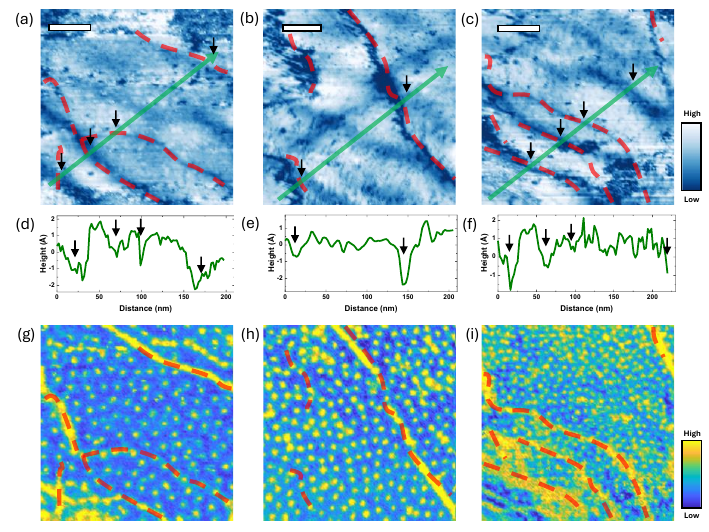}
\caption{{\bf Comparison of topography and zero bias conductance maps}. (a,b,c) STM topography corresponding to the zero bias tunneling conductance map acquired simultaneously and shown in (g-i). White bars correspond to 40 nm. The color scale corresponds to height changes of about 0.4 nm. (d-f) Height profile as a function of position along the line scan shown as a green arrow in (a-c), respectively. Red dashed lines mark the largest defects identified from the STM topography. We mark the position of linear defects with black arrows in both topography and height profiles. (g-i) The zero bias tunneling conductance map acquired simultaneously as topographies in (a-c). The color scale corresponds to conductance changes of about 0.8 times the tunneling conductance normalized to its normal state value (i.e.\,in many yellow areas, the superconducting gap is very weak but remains generally open). Red dashed lines mark the largest defects identified in the STM topography. Maps taken at 10 K and 6 T for a,g; at 10 K and 10 T for b,h; and at 15 K and 14 T for c,i.}
\label{surface}
\end{figure*}

In Supplementary Figure\,\ref{surface} we characterize the surface and its defects in three representative fields of view. The color scale following height variations along the surface in the STM topography Supplementary Figure \,\ref{surface}(a-c) has been maximized and is of the order of the atomic scale, below 0.4 nm. This is characteristic of exposed Ca or K layers\,\cite{PhysRevB.97.134501,Cao2021}. In Supplementary Figure\,\ref{surface} (d-f), we also show height profiles, marked by a green arrow in Supplementary Figure\,\ref{surface} (a-c), along several linear defects on topographies shown in Supplementary Figure\,\ref{surface}(a-c) by red dashed lines. Such linear features are often observed in CaKAs$_{4}$Fe$_4$ and related iron pnictide superconductors\,\cite{PhysRevB.97.134501,koshelev2019melting,doi:10.1126/sciadv.aat1061,wang2020,Mesaros2024}. The height changes along the linear features are below an atomic size step\,\cite{Herrera2023} and they also present sizeable pair-breaking\,\cite{PhysRevB.97.134501}. We can associate these linear defects with intercalated layers of CaFe$_2$As$_2$ and KFe$_2$As$_2$\,\cite{Ishida2019,Pyon2019,Sugali2021,Ichinose2021,Wang2021}. Some of these layers have been found to be single atomic layers (along c-axis) and tens of nm wide (in the ab-plane)\,\cite{Ishida2019,Pyon2019}. Large strain fields\,\cite{SPRINGHOLZ199712} have been observed around these single intergrown layers along the c axis (perpendicular to the layer plane)\,\cite{Sugali2021,Ichinose2021}. We expect that the corrugation observed at the surface stemming from an underlying intergrown layer is below interatomic distances, and we thus associate the linear features we observe at the surface with intergrown layers. We can find a relation between the position of these linear defects in the topography and the corresponding arrangement of vortices, as shown by the red dashed lines in the topographies  (Supplementary Figure \,\ref{surface}(a-c)) and the zero bias tunneling conductance maps (Supplementary Figure\,\ref{surface}(g-i)). Vortices are often pinned along the linear features, as shown for example in the bottom right of Supplementary Figure\,\ref{surface}(g). The maps show blurred vortices along these lines. This suggests that vortices are strongly fluctuating along the linear features. Moreover, we observe that vortex positions (Supplementary Figure\,\ref{surface}(g-i)) are oriented following these linear features. In Supplementary Figure\,\ref{Esquema pinning} we schematically present a possible arrangement of vortex positions and intergrowth of CaFe$_2$As$_2$ and KFe$_2$As$_2$.

\begin{figure}[htb]
\includegraphics[width=0.95\columnwidth]{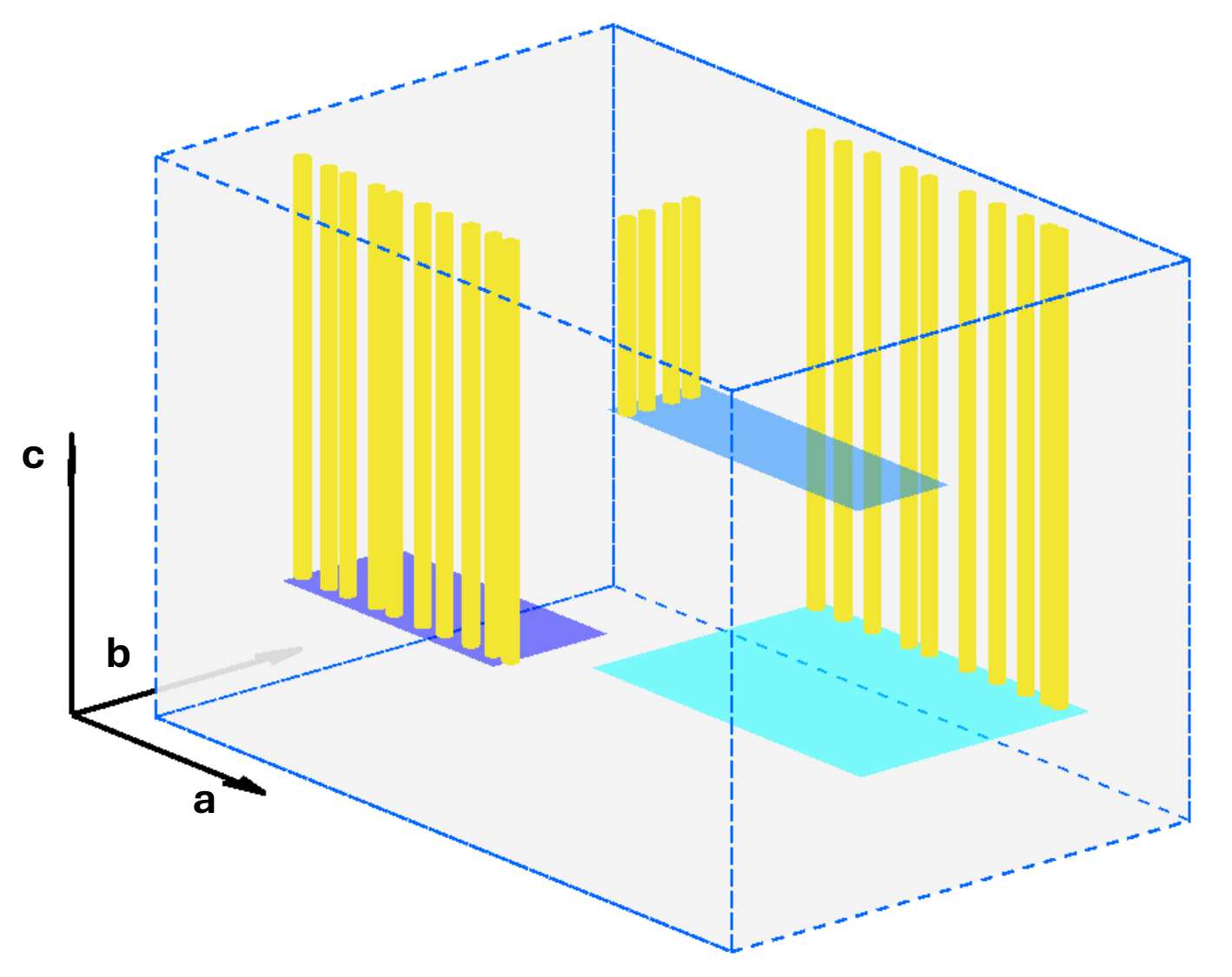}
\caption{{\bf Possible role of defects and intergrowth in vortex pinning of CaKAs$_{4}$Fe$_4$}. Schematic representation of the CaKFe$_4$As$_4$ crystal lattice with an interstitial single layers of CaFe$_2$As$_2$ or KFe$_2$As$_2$ marked with an blue planes. We represent the vortices as yellow cylinders that become pinned on the borders of these interstitial layers.}
\label{Esquema pinning}
\end{figure}

\section{Vortex core size vs temperature}

In Supplementary Figure\,\ref{cores} we show the dependence of the vortex core size as a function of the magnetic field and temperature. We observe that at low temperatures the vortex core size roughly follows the zero temperature magnetic field dependence\,\cite{Kogancore}. However, at high temperatures (yellow triangles in Supplementary Figure\,\ref{cores}), vortex jitter leads to an effective increase in the vortex core size at large magnetic fields.

\begin{figure}[htb!]
\includegraphics[width=0.95\columnwidth]{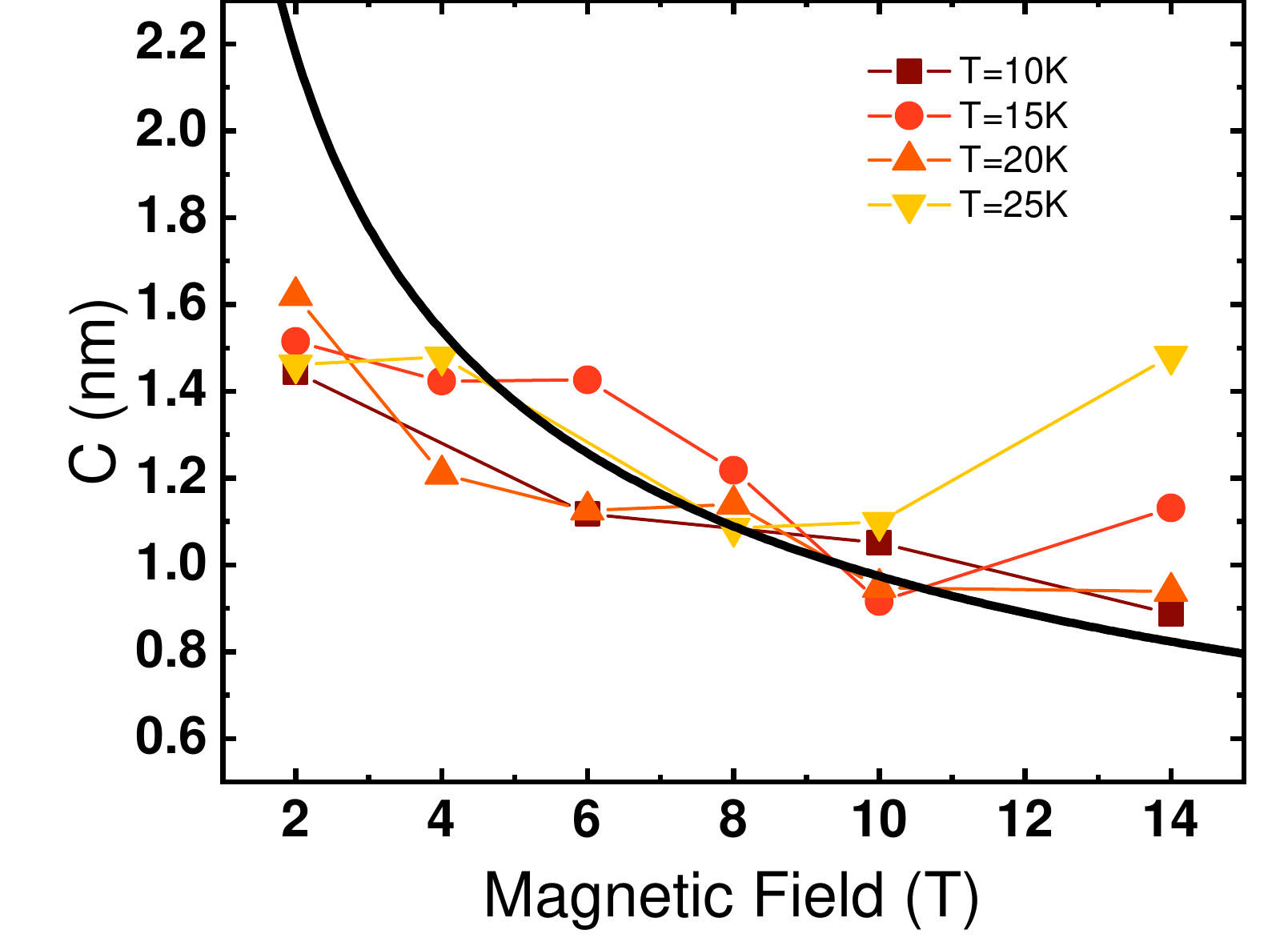}
\caption{{\bf Vortex core size, $C$ as a function of magnetic field}. The continuous black line is given by $C \propto 1/\sqrt{H} $, and is the behavior expected for the zero temperature extrapolation and found at low temperatures in CaKFe$_4$As$_4$\,\cite{Kogancore,PhysRevB.71.134505}.}
\label{cores}
\end{figure}

\section{Melting}

In Supplementary Figure\,\ref{All melting} we show examples of zero bias conductance maps at several temperatures (columns) and magnetic fields (rows). At the lowest temperatures we observe a disordered vortex solid in all cases. We also observe areas, marked by the red lines, where individual vortices are no longer resolved over a distance exceeding several times the intervortex distance. Within the same field of view, we also observe areas with a vortex solid. At 8 T the vortex solid is observed until very close to T$_c$. Just below T$_c$ we observe a vortex liquid in the full field of view. The melting temperature is at this magnetic field very close to the one obtained from macroscopic experiments.

\begin{figure*}[htb]
\includegraphics[width=1\textwidth]{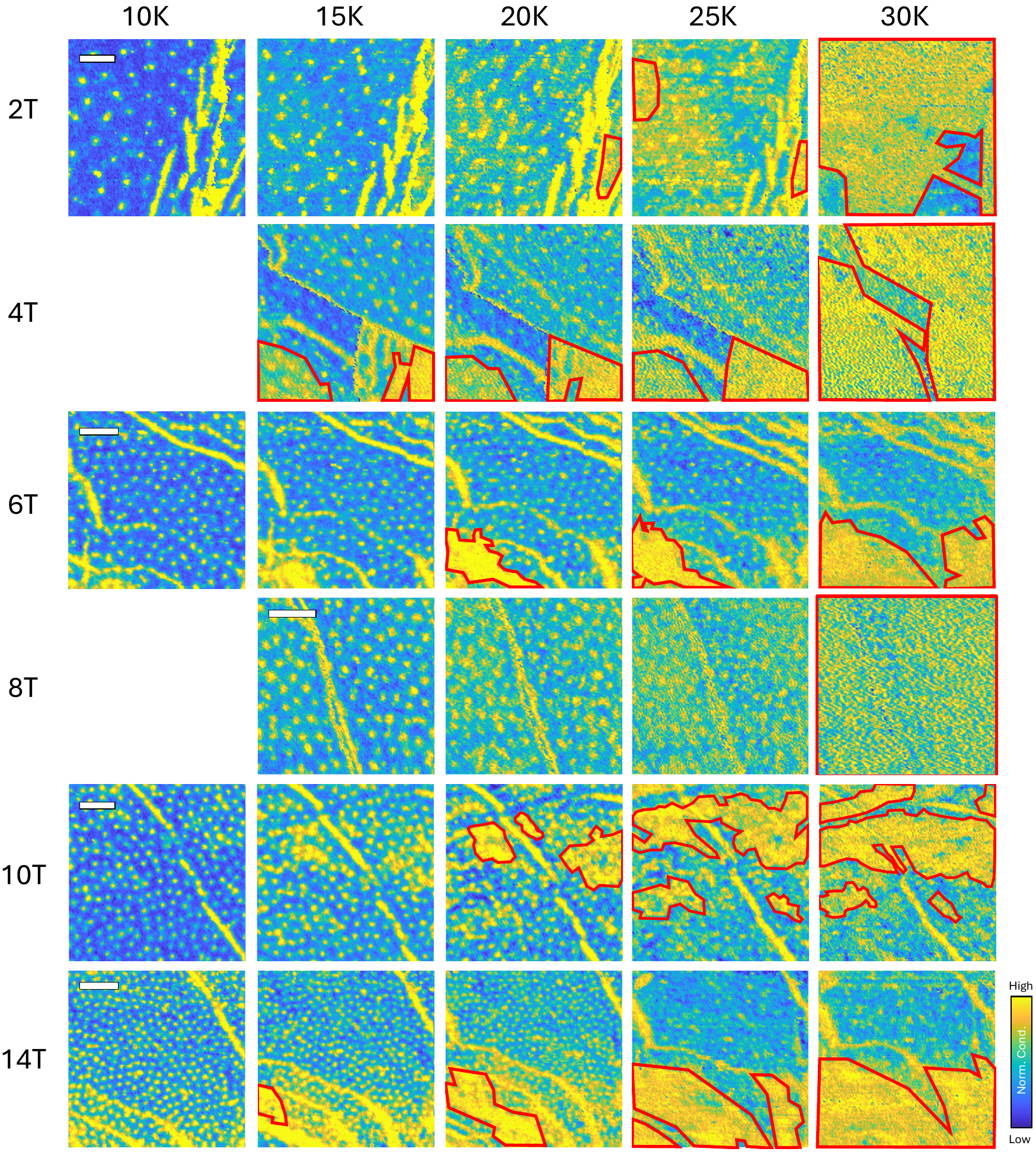}
\caption{{\bf Zero bias tunneling conductance maps as a function of the magnetic field and temperature}. Areas with vortex liquid are marked by red lines. White bars correspond to 40\,nm.}
\label{All melting}
\end{figure*}

\section{Vortex trajectories}

In Supplementary Figures\,\ref{Traj_2T}-\ref{Traj_14T} we show a few tunneling conductance maps from a series of maps taken sequentially in the same field of view, for the temperatures marked in each figure and for magnetic fields ranging from 2\,T (Supplementary Figure\,\ref{Traj_2T}) to 14\,T (Supplementary Figure\,\ref{Traj_14T}). We show half of all measured maps taken in each case. In every figure, we show in the right column the vortex positions for all frames as red dots. The red lines show the changes in the vortex positions in subsequent images. We ensured continuity by requiring displacements between successive images to remain well below the intervortex spacing. We highlight in white the vortex that presented the largest motion. We obtained the accumulative distance from the sum of the displacements of each vortex. We plot in Figure\,4(a) of the main text the accumulated distance of the vortex showing largest motion (in white in the right column of Figs.\,\ref{Traj_2T}-\ref{Traj_14T}). For low and high magnetic fields (2\,T Supplementary Figure\,\ref{Traj_2T}, 4\,T Supplementary Figure\,\ref{Traj_4T} and 14 T Supplementary Figure\,\ref{Traj_14T}) there are several vortices which move across large distances as compared to the intervortex separation. By contrast, for intermediate fields such as in 8\,T Supplementary Figure\,\ref{Traj_8T} and 10\,T Supplementary Figure\,\ref{Traj_10T} most of the vortices remain at the same position. There is no strong temperature dependence of the maximal vortex motion, in agreement with the values of the maximal accumulated distance falling onto each other for each magnetic field in Figure\,4(a) of the main text.

\begin{figure*}[htb]
\includegraphics[width=1\textwidth]{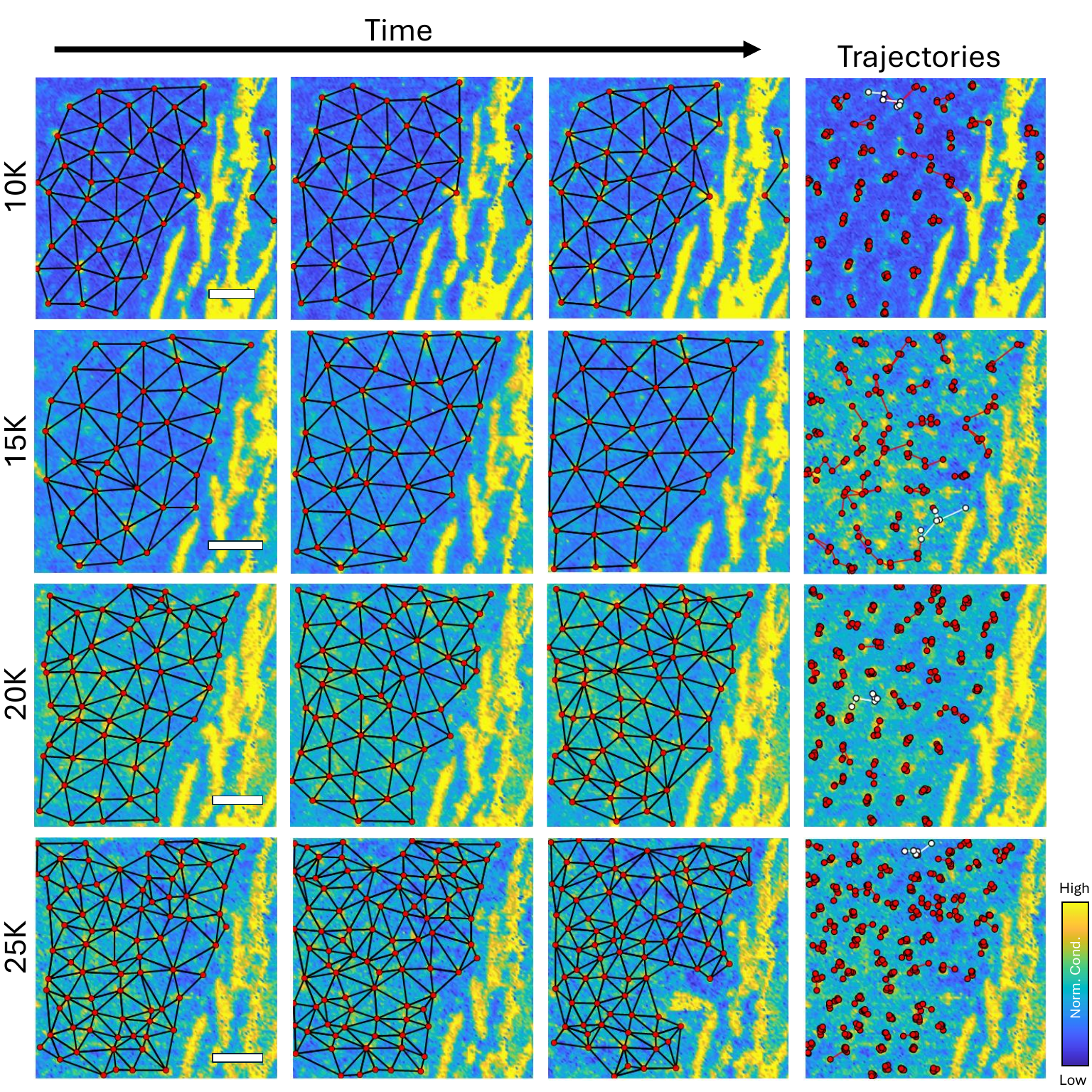}
\caption{{\bf Zero bias tunneling conductance maps at 2\,T.} Each row corresponds to images taken at different temperatures (10\,K-25\,K) in the same field of view. The first three columns show three representative images at the start (left), middle (middle) and end (right) of the sequence. The red dots mark the positions of the vortices, and the black lines connecting them are the triangulation of the vortex positions. The last column shows the extracted trajectories, with red lines joining the vortex positions in different frames. The vortex highlighted in white is the one that shows the largest accumulated distance. White bars correspond to 40\,nm.}
\label{Traj_2T}
\end{figure*}

\begin{figure*}[htb]
\includegraphics[width=1\textwidth]{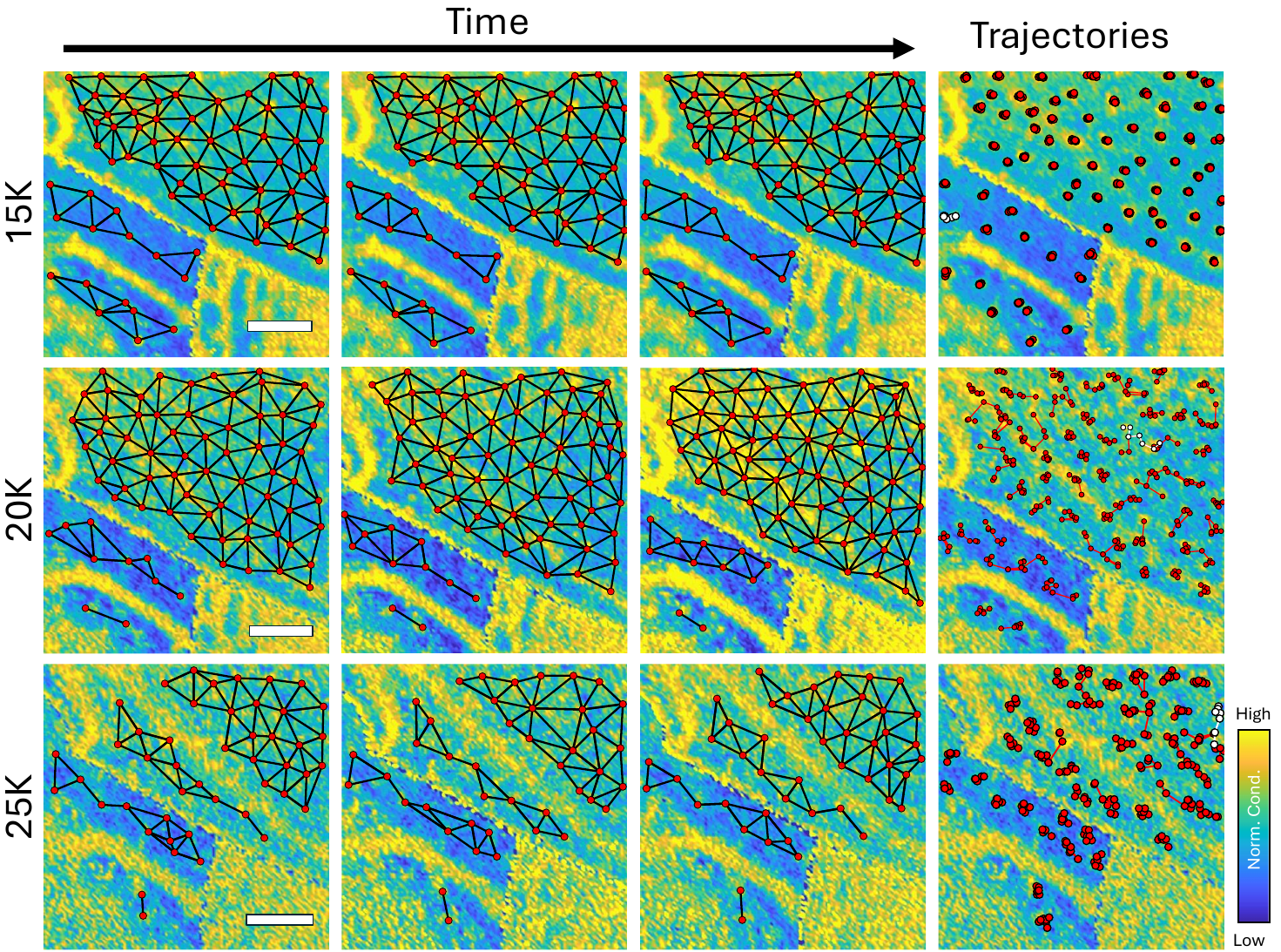}
\caption{{\bf Zero bias tunneling conductance maps at 4\,T}. Each row corresponds to images taken at different temperatures (15\,K-25\,K) in the same field of view. The first three columns show three representative images at the start (left), middle (middle) and end (right) of the sequence. The red dots mark the positions of the vortices, and the black lines connecting them are the triangulation of the vortex positions. The last column shows the extracted trajectories, with red lines joining the vortex positions in different frames. The vortex highlighted in white is the one that shows the largest accumulated distance. White bars correspond to 40\,nm.}
\label{Traj_4T}
\end{figure*}

\begin{figure*}[htb]
\includegraphics[width=1\textwidth]{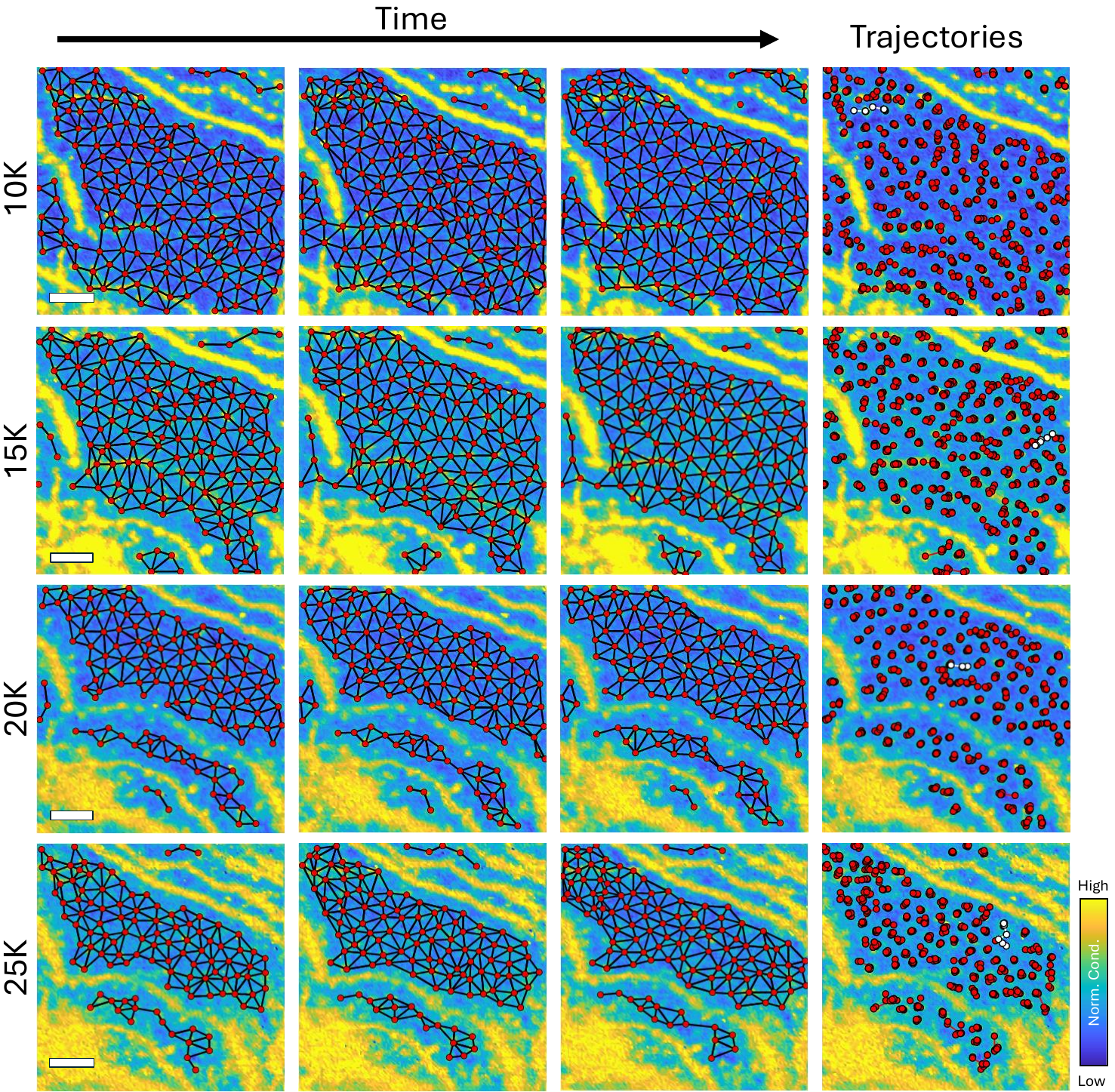}
\caption{{\bf Zero bias tunneling conductance maps at 6\,T}.  Each row corresponds to images taken at different temperatures (10\,K-25\,K) in the same field of view. The first three columns show three representative images at the start (left), middle (middle) and end (right) of the sequence. The red dots mark the positions of the vortices, and the black lines connecting them are the triangulation of the vortex positions. The last column shows the extracted trajectories, with red lines joining the vortex positions in different frames. The vortex highlighted in white is the one that shows the largest accumulated distance. White bars correspond to 40\,nm.}
\label{Traj_6T}
\end{figure*}

\begin{figure*}[htb]
\includegraphics[width=1\textwidth]{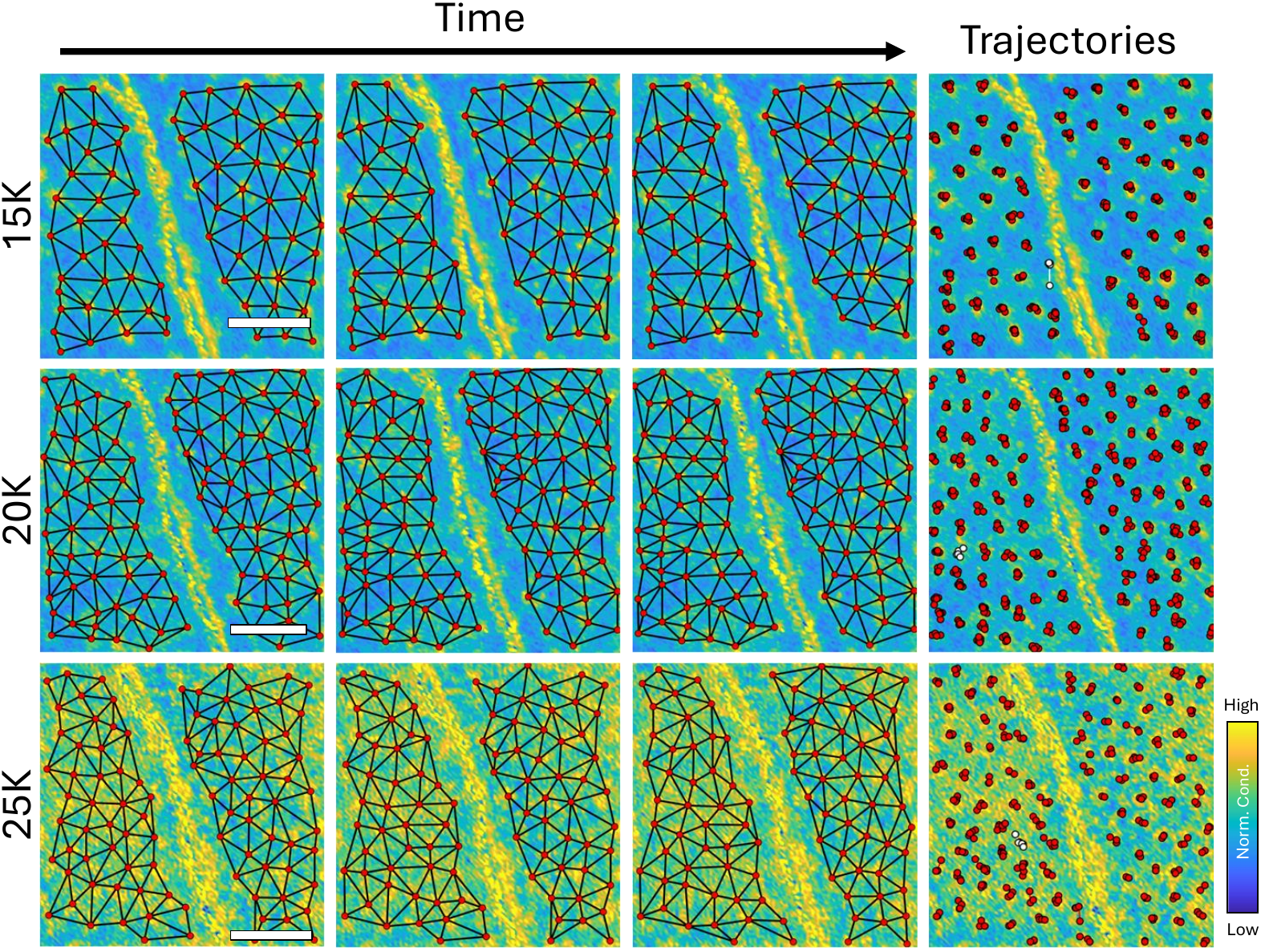}
\caption{{\bf Zero bias tunneling conductance maps at 8\,T}.  Each row corresponds to images taken at different temperatures (15\,K-25\,K) in the same field of view. The first three columns show three representative images at the start (left), middle (middle) and end (right) of the sequence. The red dots mark the positions of the vortices, and the black lines connecting them are the triangulation of the vortex positions. The last column shows the extracted trajectories, with red lines joining the vortex positions in different frames. The vortex highlighted in white is the one that shows the largest accumulated distance. White bars correspond to 40\,nm.}
\label{Traj_8T}
\end{figure*}

\begin{figure*}[htb]
\includegraphics[width=1\textwidth]{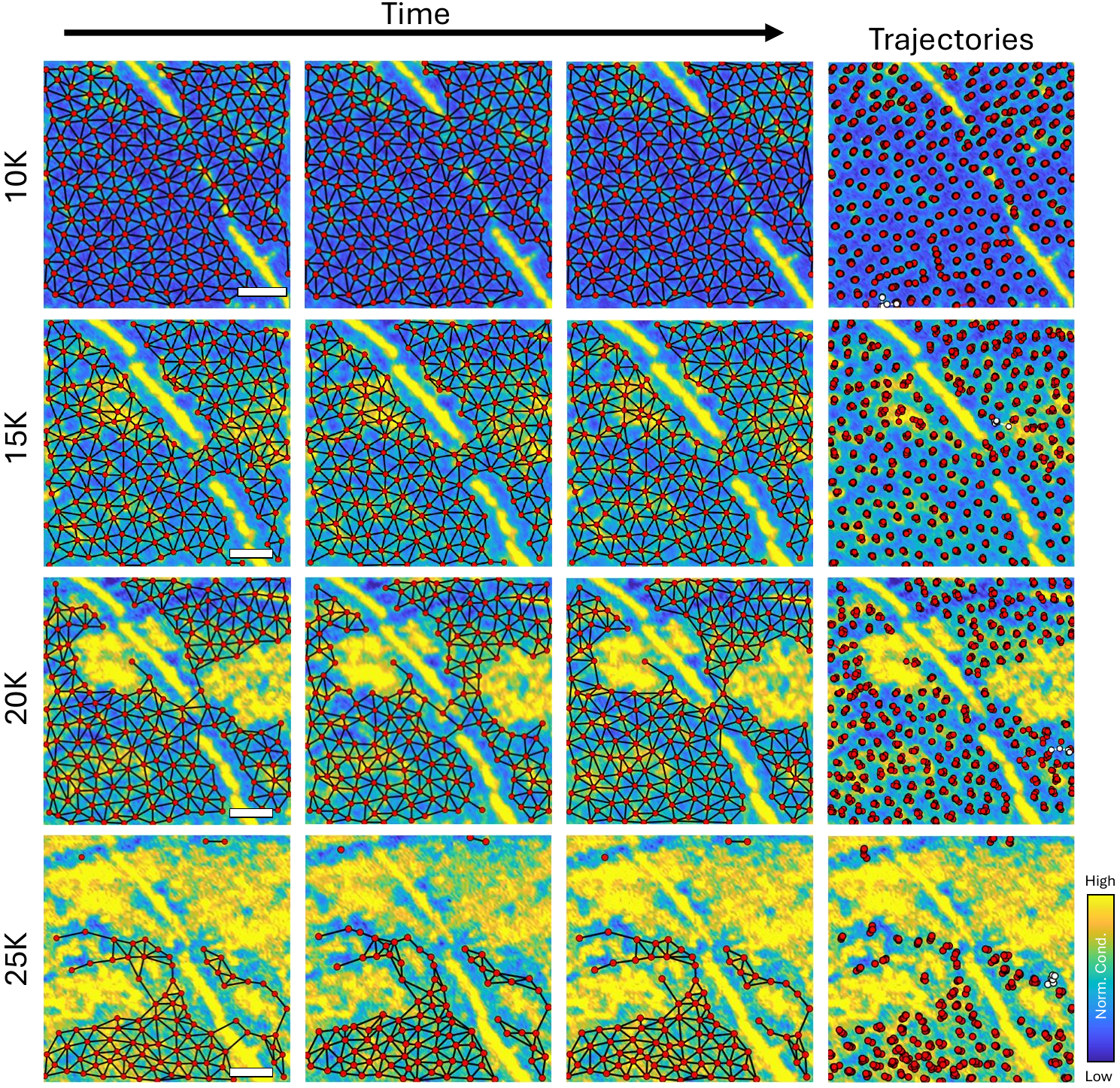}
\caption{{\bf Zero bias tunneling conductance maps at 10\,T}. Each row corresponds to images taken at different temperatures (10\,K-25\,K) in the same field of view. The first three columns show three representative images at the start (left), middle (middle) and end (right) of the sequence. The red dots mark the positions of the vortices, and the black lines connecting them are the triangulation of the vortex positions. The last column shows the extracted trajectories, with red lines joining the vortex positions in different frames. The vortex highlighted in white is the one that shows the largest accumulated distance. White bars correspond to 40\,nm.}
\label{Traj_10T}
\end{figure*}

\begin{figure*}[htb]
\includegraphics[width=1\textwidth]{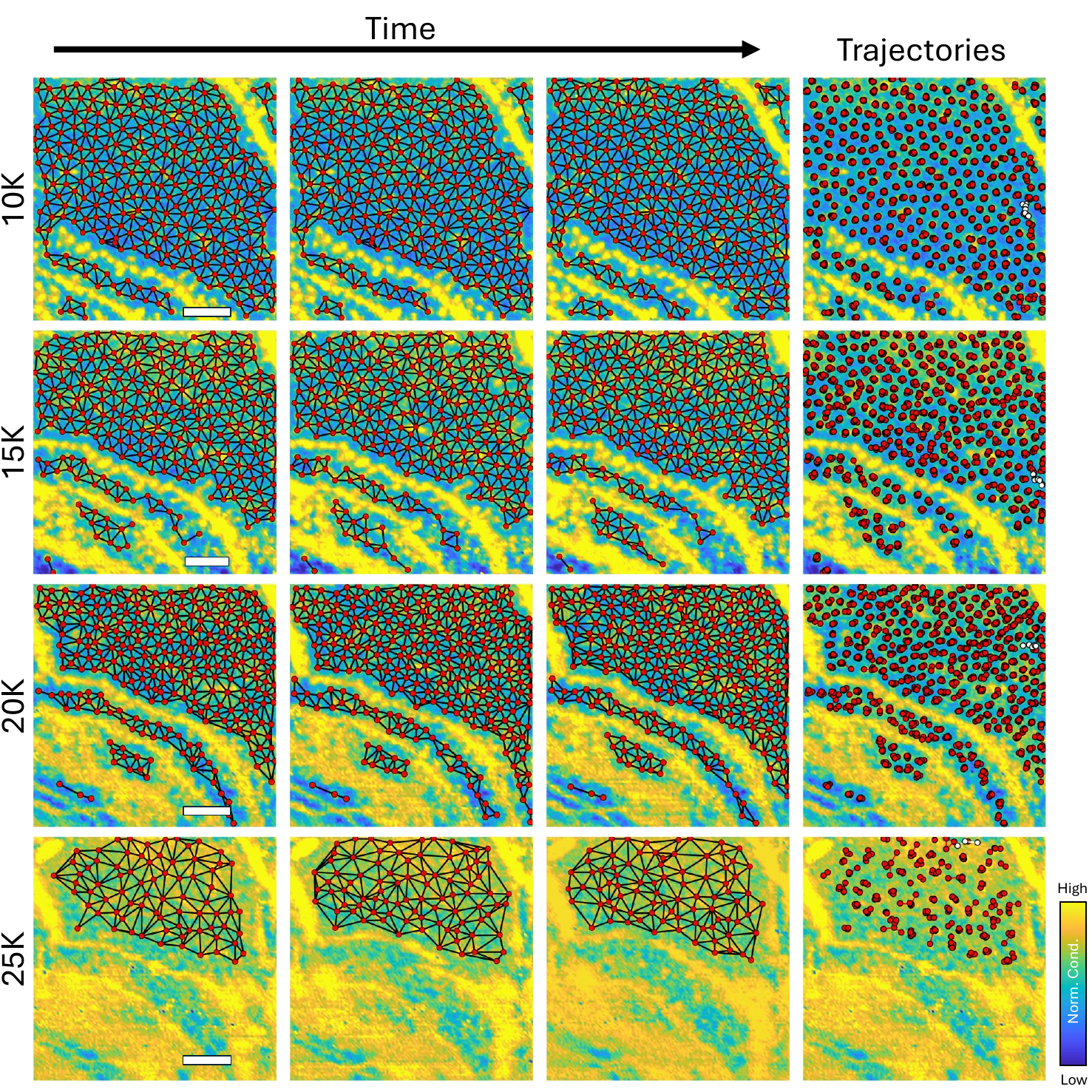}
\caption{{\bf Zero bias tunneling conductance maps at 14\,T}.Each row corresponds to images taken at different temperatures (10\,K-25\,K) in the same field of view. The first three columns show three representative images at the start (left), middle (middle) and end (right) of the sequence. The red dots mark the positions of the vortices, and the black lines connecting them are the triangulation of the vortex positions. The last column shows the extracted trajectories, with red lines joining the vortex positions in different frames. The vortex highlighted in white is the one that shows the largest accumulated distance. White bars correspond to 40\,nm.}
\label{Traj_14T}
\end{figure*}

\section{Separation of vortices}
 
Changes in the vortex positions due to thermal fluctuations (Figure\,1(b) of the main text) lead to an increase in the zero bias conductance around the vortex core with respect to the absence of fluctuations (Figure\,1(a) of the main text). The time scale involved in the fluctuations depends on the temperature and the pinning landscape and mostly involves rapid fluctuations, as schematically shown in Figure\,1 of the main text. However, when vortices are in between close-lying pinning centers, fluctuations might considerably slow\,\cite{PhysRevB.105.144504,PhysRevResearch.2.043266}. We discuss an extreme example of this phenomenon now.

In Supplementary Figure\,\ref{Separation} we present a zoom to a certain area at 10\,T and 10\,K. We mark in Supplementary Figure\,\ref{Separation}(a-b) (left panels) one vortex by a dashed circle. Apparently, this is a single vortex. However, tracing the conductance profile as a function of the position through the apparent vortex center leads to a two-peak structure (right panels in Supplementary Figure\,\ref{Separation}). We can fit the profile using two Gaussians located approximately 10\,nm apart. The average intervortex distance at 10\,T is of about 15\,nm and the vortex core size, see Section II, is below 2\,nm. This suggests that the feature observed in the zero bias tunneling conductance marked by the white dashed circle in Supplementary Figure\,\ref{Separation} is composed of two vortices. Indeed, with time, this feature separates into two features, as shown by the white dashed ellipses in the left panels of Figs.\,\ref{Separation}(c-f). The vortex profiles (shown in the right panels of Figs.\,\ref{Separation}(c-f)) consist now of two well-separated Gaussian shapes. Their separation reaches the intervortex distance for this magnetic field at the end of the sequence (Supplementary Figure\,\ref{Separation}(f)). 

This slowly fluctuating vortex is an extreme example of vortex fluctuations on long time scales and occurs in an area where the rest of the vortices are well pinned. The areas where we observe vortex droplets (red areas in Figure\,2\,(c-e) of the main text) generally coincide with areas where subsequently acquired tunneling conductance maps present vortices fluctuating slightly around a certain position, for example at the top left and right parts of the tunneling conductance map of Figure\,2(b) of the main text.

\begin{figure*}[htb]
\includegraphics[width=1\textwidth]{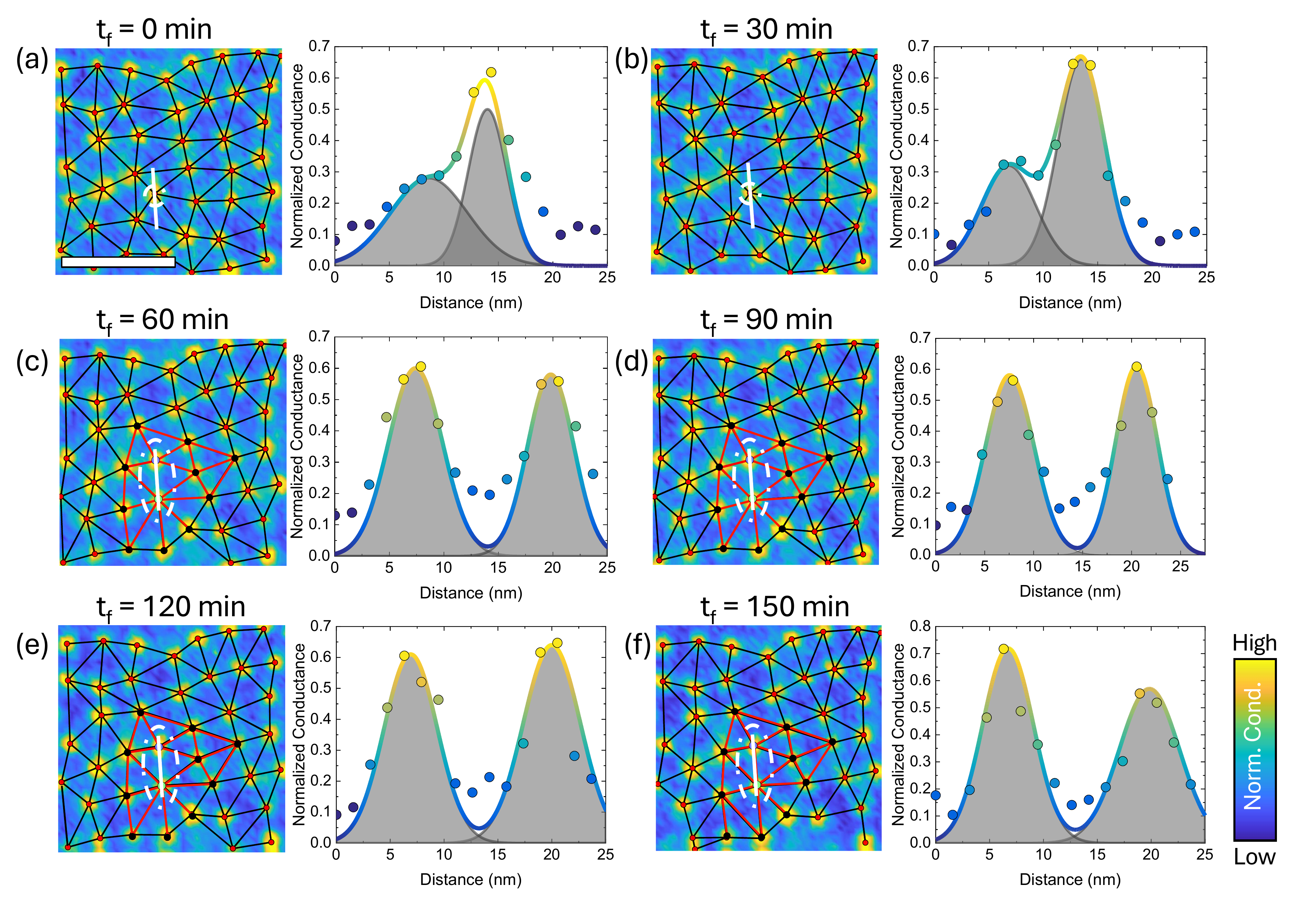}
\caption{{\bf Vortex fluctuations in a small field of view.} In panels (a–f) we focus on the portion of the 10\,T maps showing an extreme example of slow vortex creep, presenting subsequently acquired maps from (a) to (f). We show again the zero bias tunneling conductance maps, together with vortex positions (red circles) and nearest neighbor structure (black lines) in the left panels. Red lines connecting the red circles correspond to the lines of the triangulation modified by the splitting of an apparent single vortex in (c). In the right panels, we show profiles of the tunneling conductance as dots and fits to Gaussians as lines. Each Gaussian is shown separately in grey below the fit. We mark with a white dashed circle in the left panels an area which seems to contain a single vortex in (a), which splits into two after 150 minutes in (f). White bar in (a) correspond to 40 nm.}
\label{Separation}
\end{figure*}

\clearpage

\bibliographystyle{naturemag}

\end{document}